\documentclass{tlp}  

\usepackage{times}

\usepackage{xspace}
\usepackage{url}
\usepackage{graphicx} 
\usepackage{epstopdf}
\usepackage{color}

\newcommand{\notf}[1]{\ensuremath{\mathtt{not\ } #1}}

\usepackage{microtype}
 
\newcommand{\aspdocBf}{\sc{\bfseries{{ASPDoc}}}\xspace}
\newcommand{\aspunitBf}{{\sc{\bfseries{ASPUnit}}\xspace}}
\newcommand{\lang}{{\sc{Lana}\xspace}}
\newcommand{\langbf}{{\sc{Lana}\xspace}}
\newcommand{\aspdoc}{{\sc{ASPDoc}}\xspace}
\newcommand{\javadoc}{{\sc{JavaDoc}}\xspace}
\newcommand{\aspunit}{{\sc{ASPUnit}}\xspace}
\newcommand{\plunit}{{\sc{PLUnit}}\xspace}
\newcommand{\junit}{{\sc{JUnit}}\xspace}

\newcommand{\sealion}{\ensuremath{\mathtt{SeaLion}}\xspace}
\newcommand{\aspide}{\ensuremath{\mathtt{Aspide}}\xspace}
\newcommand{\DLV}{\ensuremath{\mathtt{DLV}}\xspace}
\newcommand{\gringo}{\ensuremath{\mathtt{gringo}}\xspace}
\newcommand{\clasp}{\ensuremath{\mathtt{clasp}}\xspace}
\newcommand{\clingo}{\ensuremath{\mathtt{clingo}}\xspace}
\newcommand{\ape}{\ensuremath{\mathtt{APE}}\xspace}

\newcommand{\egc}{e.g.,\xspace}
\newcommand{\iec}{i.e.,\xspace}
\newcommand{\wrt}{with respect to\xspace}

\newcommand{\nonterm}[1]{\emph{#1}}
\newcommand{\term}[1]{``{\tt#1}''}

\submitted{25 March 2012}
\revised{11 June 2012}
\accepted{18 June 2012} 
 
\begin{document}

\title[Annotating Answer-Set Programs in \langbf]{Annotating Answer-Set Programs in \langbf\footnote{%
This work was partially supported by the Austrian Science Fund (FWF) under grant P21698. 
}}

\author[Marina De Vos, Do\u{g}a K{\i}za, Johannes Oetsch, J\"org P\"{u}hrer, Hans Tompits]{
MARINA DE VOS \\
Department of Computing\\
University of Bath, BA2 7AY,
Bath, UK   \\
\email{mdv@cs.bath.ac.uk}
\and
DO\u{G}A GIZEM 
KISA\\
Faculty of Engineering and Natural Sciences, \\
Sabanci University, Orhanli, Tuzla, Istanbul 34956, Turkey \\
\email{dgkisa@su.sabanciuniv.edu}
\and
JOHANNES OETSCH, J\"ORG P\"UHRER, HANS TOMPITS\\
Institut f\"ur Informationssysteme 184/3, \\
Technische Universit\"at Wien, Favoritenstra{\ss}e 9-11, A-1040 Vienna, Austria \\
\email{\{oetsch,puehrer,tompits\}@kr.tuwien.ac.at} 
}

\pagerange{\pageref{firstpage}--\pageref{lastpage}}
\volume{\textbf{10} (3):}
\jdate{January 2012}
\setcounter{page}{1}
\pubyear{2012}

\maketitle
\label{firstpage}

\begin{abstract}
While past research in answer-set programming (ASP) mainly focused on theory, ASP solver technology, and applications,
the present work situates itself in the context of a quite recent research trend:  \emph{development support for ASP.}
In particular, we propose to augment answer-set programs with additional meta-information formulated in a dedicated annotation
language, called \lang. This language allows the grouping of rules into coherent blocks and  to specify language signatures, types, pre- and postconditions, as well as unit tests for such blocks. While these annotations are invisible to an ASP solver, as they take the form  of program comments, they can be interpreted by tools for documentation,  testing, and verification purposes, as well as to eliminate sources of common programming
errors by realising syntax checking or code completion features. To demonstrate its versatility, we introduce 
two such tools, viz.\ (i)~\aspdoc, for 
generating an HTML documentation for a program based on the annotated information, and (ii)~\aspunit, for running and monitoring unit tests on program blocks. \lang\ is also exploited in the \sealion system, an integrated development environment for ASP based on Eclipse.
\end{abstract}

\section{Introduction}\label{sec:intro}

Answer-set programming (ASP)~\cite{gellif88,gellif91,baral03} is an established approach  for declarative problem solving and non-monotonic reasoning.
So far,  research in ASP can basically be classified into three  categories:
(i)
theoretical foundations of ASP including language extensions,
(ii)
performance and capabilities of ASP solvers, and
(iii)
case studies and applications involving ASP. 
More recently, methods and methodologies to support an ASP programmer are increasingly becoming a 
focus of research interest~\cite{sea07,sea09,oetsch10}.

In this paper, we propose to augment programs with additional meta-information to facilitate the ASP development process.  To this end, 
we devised  a dedicated annotation language, {\lang}, standing for ``{L}anguage for {AN}notating {A}nswer-set programs'',
that specifies specially formatted program comments.  This meta-information 
is invisible to an ASP solver---therefore not altering the semantics of the program---but different tools may interpret and use the annotated information to various ends like documentation, testing, verification, code completion, or other development support.

One particular and quite central feature of \lang\ is grouping rules that
are related in meaning into coherent blocks.
Although different notions of modularity have already been 
proposed for ASP in the literature~\cite{bugliesi94,eiter97,gelfond99,balduccini07,janhunen09}, a strict concept of a program module can sometimes
be a too tight corset.  
 In particular, notions of modularity in ASP often come with their own semantics and different kinds of constraints need to
be satisfied. For example, DLP-functions~\cite{janhunen09} require that their input and output signatures are disjoint and two DLP-functions
need to satisfy certain syntactic constraints in order to be composable.
On the other hand, $lp$-functions are a modular approach to build a logic program from its specification~\cite{gelfond99}. That is, an $lp$-function
is used to realise some functional specification that relates input and output relations for some domain by means of a logic program. 
The kind of grouping that we are proposing has, however, 
no semantical ramifications other than documenting that some rules belong together in a certain sense. 
Nevertheless, the benefit is that we add some extra structure to a program, improving 
the clarity and coherence of the program  parts, which in turn can be used by other tools for, \egc unit testing~\cite{beck03}.
While unit testing is an integral element in software development using common languages like Java or C, it has
been addressed in ASP only quite recently~\cite{febbraro11}. We provide  means to formulate unit tests for single blocks using \lang, allowing for easy regression testing.
After rules are grouped into blocks, we may use further annotations to declare respective input and output signatures, which are also useful for testing and verification. Furthermore, we can declare the names and arities of predicates that are used within a block. This information
can be exploited by, \egc an integrated development environment (IDE) for syntax checking and code completion features while a user
is writing a program.  Besides names, description, and arities of predicates, one can also specify the domains of term arguments of a predicate using
respective language features for declaring types. This information can be used for automated detection of type violations. These declarations 
have the potential to
eliminate the source for quite common programmer errors with only little extra cost.
For verification purposes, our annotation language can be used to specify assertions like pre- and postconditions for blocks. Preconditions are expected to
hold for any input of a block, while postconditions have to hold for any output. Together, they can be used to verify correctness of an ASP encoding \wrt
such assertions.

Our main contributions are thus as follows:
\begin{itemize}
\item We introduce an annotation language for ASP that offers various ways to  express meta-information for rules and other language elements. 
This information can be used to support and ease the development process, test and verify programs, and to eliminate many sources for 
common programmer errors.
\item We describe \aspdoc, a \javadoc\footnote{\url{http://www.oracle.com/ technetwork/java/javase/documentation/index-jsp-135444.html}.} 
 inspired tool, which takes an answer-set program, interprets the meta-comments, and automatically generates 
an HTML file documenting the program.
\item We introduce \aspunit, an implementation of a unit-testing framework in the spirit of \junit\footnote{\url{http://www.junit.org}.} 
 based on the structural annotations found in a program. 
This framework allows to
formulate unit  tests for individual program blocks, to execute them, and to monitor test runs. 
\end{itemize}

The paper is organised as follows.
In Section~\ref{sec:bg}, we provide some background on ASP.
Then, in Section~\ref{sec:lang}, we introduce \lang, explain the basic language features, 
and illustrate them using a running example.
Section~\ref{sec:aspdoc} describes the basic features of \aspdoc while Section~\ref{sec:aspunit} does the same for \aspunit.
Finally, we review related work in Section~\ref{sec:rel} and conclude in Section~\ref{sec:concl} with pointers for future work.

\section{Background}\label{sec:bg}

Answer-set programming (ASP) \cite{gellif88,gellif91,baral03} is a
declarative programming paradigm in which a logic program is
used to describe the requirements that must be fulfilled by the
solutions of a certain problem.  The solutions of the problem can be 
obtained through the interpretation of the answer sets of the program,
usually defined through a variant or extension of 
 the stable-model semantics \cite{gellif88}.
This technique has been successfully applied in various domains such
as planning \cite{eiter2002,lif2000}, configuration and verification
\cite{soininen99}, music composition \cite{bobrdvfftplph}, 
or reasoning about biological networks \cite{grscse06b} to just name a few.
In the following, we briefly cover the essential concepts of ASP; for in-depth coverage, we refer to the well-known textbook by \citeN{baral03}. 

The basic components of the language are \emph{atoms}, elements that can be
assigned a truth value. An atom $a$ can be negated using {\em negation as failure}. A {\em literal} is an atom $a$ or a negated atom $\notf{a}$. We say that $\notf{a}$ is true if
we cannot find evidence supporting the truth of $a$.
Atoms and literals are used to create rules of the general form
\[a \verb! :- !  b_{1}, \ldots, b_{m}, \notf{c_{1}}, \ldots, \notf{c_{n}}\,,\] 
where $a$, $b_{i}$, and $c_{j}$ are atoms. Intuitively, this means
{\it if all atoms $b_{i}$ are known/true and no atom $c_{i}$ is known/true, then $a$ must be known/true}. We refer to $a$ as the head and
$b_{1},\ldots, b_{m}, \notf{c_{1}},\ldots, \notf{c_{n}}$ as the body of the rule. Rules with empty body are  called {\em facts}. Rules with empty head are referred to 
as {\em constraints}, indicating that no solution satisfies the body.
A (\emph{normal}) \emph{program} is a 
set of rules.
The semantics 
is defined in terms of {\em answer sets}, 
i.e., assignments of true and false to all atoms in the program 
that satisfy the rules in a
minimal and consistent fashion.
A program has zero or more answer sets,
each corresponding to a solution.

Different extensions to the language have been proposed. 
On the one hand, we have syntactic extensions, providing mere, but very useful, syntactic sugar. On the other hand, we have semantic extensions where the formalism itself is generalised.

From a programmer's perspective, \emph{choice rules}~\cite{NiemelaSS99} are probably the most commonly used extension.
Many problems require choices between a set of atoms to be made. Although this can be modelled in the basic formalism,
it tends to increase to the number of rules and increases the possibility of errors. To avoid this, choices are introduced. Choices, written $L \{l_1, \ldots, l_n\} M$, are a convenient construct to indicate that at least $L$ and at most $M$ literals from the set $\{l_1, \ldots, l_n\}$ must be true in order to satisfy the construct.  Bound $L$ defaults to 0 while $M$ defaults to $n$. Choice rules are often used  with a grounding predicate: $L \{A(X):B(X)\} M$ represents the choice of a number of atoms  where $A(X)$ is grounded with all values of $X$ for which $B(X)$ is true.

One of the major extensions to the language was the introduction of disjunction in the head of rules~\cite{gellif91}.
When the body of the rule is true, we need to have at least one head atom that is true. 
Although at first it may seem that disjunctive programs can easily be translated to non-disjunctive programs, this is not the case. Both types of programs are in different complexity classes  (under the usual complexity-theoretic proviso that the polynomial hierarchy does not collapse). 

Algorithms and implementations for obtaining answer sets of logic
programs are referred to as {\em answer-set solvers}.  The most
popular and widely used solvers are \textsc{DLV}~\cite{eilemapfsc98}, providing solver capabilities for
disjunctive programs, and the SAT inspired \textsc{clasp} \cite{gekanesc07a}. Alternatives are 
\textsc{Smodels}~\cite{nisi97} and \textsc{Cmodels}~\cite{sat_asp}, a solver based on translating the
program to SAT.

\section{An Annotation Language for ASP}\label{sec:lang}
\begin{figure}[t]
\figrule
\begin{center}
\includegraphics[width=0.3\textwidth]{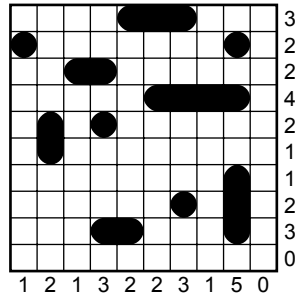}
\end{center}
\figrule
\caption{Solved instance of a Battleship puzzle.}
\label{fig:battleship}
 \end{figure}

In this section, we describe our annotation language \lang. 
 An overview of most language elements of \lang\ appears in Table~\ref{tab:summary}; Table \ref{tab:testcases} gives a summary of the remaining language
 elements related to testing in \lang.
We make use of a simple answer-set program to illustrate all the
language features in a step-by-step fashion.
In particular, we use an encoding of the \emph{Battleship} puzzle. 
A solved instance of
Battleship appears in Fig.~\ref{fig:battleship}. 
In Battleship, a group of ships is hidden on a grid and one has to find the positions of each. 
The fleet consists of one four-squares long battleship, two three-squares long cruisers,  three two-squares long destroyers, and four one-square long submarines---each ship is placed horizontally or vertically on the 
grid such that no ship is touching any other ship (not even diagonally). 
To provide hints, some squares may show parts of a ship or water. Moreover, a number besides each row and each column indicates how many
squares in that row or column are occupied by ship parts. 
A solution of a puzzle is a configuration of all the ships that is consistent with the initially given hints.

\begin{table}
\caption{Overview of \lang\  based on  BNF. 
}\label{tab:summary}
\vspace{-0.5\baselineskip}
\centering\small
\begin{tabular}{cll}
\hline\hline
Element  &  Definition & Informal Description\\
\hline
\nonterm{block } & \nonterm{block} $\mid$ \nonterm{atom} $\mid$ \nonterm{term} $\mid$  \nonterm{input}  & \lang\ elements related to blocks.   \\
\nonterm{element}      & \nonterm{signature} $\mid$  \nonterm{output signature} $\mid$                &  \\
       &  \nonterm{precondition}  $\mid$ \nonterm{postcondition}  &  \\      
\hline
\nonterm{block} & \term{@block} \nonterm{name} \term{\{} & Groups ASP rules into coherent parts. \\
       &    [\nonterm{description}] 	\{\nonterm{block element}\}	&  \\
       &   [\nonterm{ASP code}]  \term{\}} &  \\
\hline
\nonterm{atom}      &	\term{@atom} \nonterm{name} \term{(}\nonterm{termList}\term{)}	& Defines a predicate; \nonterm{termList} are the\\
					&  	[\nonterm{description}] 					&   predicate's arguments. \\ 
\hline
\nonterm{term}      & \term{@term} \nonterm{name}  & Declares a term from some \nonterm{atom} term \\ 
       		&    [\nonterm{description}] 	[\nonterm{type}] &  list, its meaning, and type information. \\       
     
\hline

\nonterm{type}	& \term{@from} \nonterm{groundTerms} $\mid$ & Type of a term is defined  by a list of  \\ 
							& \term{@with} \nonterm{ruleBdy} $\mid$		& ground terms, the  terms satisfying   \\ 
							& \term{@samerangeas} \nonterm{term} 	& \nonterm{ruleBdy}, or as the type of another term.\\

\hline
\nonterm{input }     & \term{@input} \nonterm{inputPredicates} $\mid$& Declares  input predicates of a block as a   \\
\nonterm{signature}	 &  \term{@requires} \nonterm{inputPredicates} 									& list of name/arity pairs.	\\	
\hline
\nonterm{output }    & \term{@output} \nonterm{outputPredicates} $\mid$ &  Declares  output predicates of a block as a  \\
\nonterm{signature}	&  \term{@defines} \nonterm{outputPredicates}  &list of name/arity pairs. \\	
\hline
\nonterm{assertion}    & \term{@assert}  \nonterm{name}  \term{\{}  & A logical condition for answer sets. \\
	& [\nonterm{description}] \nonterm{assertspec} \term{\}}  &  \\
\hline
\nonterm{pre-}    & \term{@precon} \nonterm{name} \term{\{} & A logical condition for the inputs of a\\
\nonterm{condition}								& 
[\nonterm{description}] \nonterm{assertSpec} \term{\}} & block.\\
\hline
\nonterm{post-}    & \term{@postcon} \nonterm{name}  \term{\{}   & A logical condition for the answer sets \\
\nonterm{condition}								& [\nonterm{description}] \nonterm{assertspec} \term{\}}  & of a block. \\
\hline
\nonterm{assertspec}    & (\term{@always} $\mid$ \term{@never}) \nonterm{atmList}	& The testmode for assertions, preconditions\\								&  
[\nonterm{embASPcode}]
	&   and postconditions; \nonterm{embASPcode} is  code\\
&   	&   within the \lang\ comment environment. \\
\hline
\hline
\end{tabular}
\vspace{-1.2\baselineskip}

\end{table}

\begin{table}
\caption{Test case specification in \lang. 
}\label{tab:testcases}
\vspace{-0.5\baselineskip}
\centering\small
\begin{tabular}{cll}
\hline\hline
Element  &  Definition & Informal Description\\
\hline

\nonterm{test case} & \term{@testcase} \nonterm{name} & A test case for the blocks from \nonterm{blocks}\\
					& [\nonterm{description}]	 \term{@scope} \nonterm{blocks} 					 & passes if the blocks joined  with \nonterm{ASP code} \\
					&\nonterm{testcond}	[\nonterm{ASP code}]				& satisfy all specified test conditions.\\
\hline
\nonterm{testcond}	&	\term{@testhasanswerset} $\mid$		& A test condition holds if one, resp., no,   \\ 
							& \term{@testnoanswerset} $\mid$		&   answer   set is found, or all ground atoms	\\ 
							& \term{@testatoms} \nonterm{atmList} \nonterm{mode}	& in \nonterm{atmList} are entailed according to \nonterm{mode}.  \\
\hline
\nonterm{mode}	&	\term{@trueinall} $\mid$							&	Mode of entailment of a test condition,  \\ 
							& \term{@trueinatleast} $n$ $\mid$	&   \iec if test atoms are true, resp., false, in 	\\ 
							& \term{@trueinatmost}	$n$ $\mid$	& all, at most $n$, or at least $n$ answer-sets. \\ 
							& \term{@falseinall}	$\mid$						&	\\ 
							& \term{@falseinatleast} $n$ $\mid$	& \\ 
							& \term{@falseinatmost} $n$	& \\
\hline
\hline
\end{tabular}
\vspace{-1.2\baselineskip}

\end{table}

Assume that a puzzle instance is defined in terms of facts \verb!water/2! and \verb!ship/2!
for specifying which squares contain water or parts of a ship, respectively.
The facts \verb!rowHint/2! and \verb!colHint/2! determine the numbers associated with each row and each column.
Problem solutions are represented by facts \verb!ship(W,X,Y,Z)! expressing that a ship is occupying the squares from
position \verb!(W,X)! to \verb!(Y,Z)!.

\subsection{Blocks}
 
In general, all keywords of our annotation language start with the symbol \verb!@!.
A central feature of \lang\ is to group rules together. This is done
using the \verb!@block! keyword.
To specify that we are going to define a couple of rules that
encode solutions to the Battleship puzzle, we declare a block with the name \verb!Battleship! as follows:

{\small
\begin{verbatim}
%** @block Battleship { *%

 % encoding of the Battleship puzzle

%** } *%
\end{verbatim}
}

The annotation \verb!@block! is followed by an optional name of the block and the opening bracket ``\verb!{!''. Everything between
``\verb!{!'' and the closing bracket  ``\verb!}!'' now belongs to the block \verb!Battleship!. 
Blocks can be nested but they must not overlap.

Note that every annotation has the form of an ASP comment. 
ASP block-comments can be used instead of starting every single line with ``\verb!%!'' 
when an ASP solver, in particular its grounding component,  supports this feature. 
To distinguish annotations from ordinary (block-)comments, an additional star ``\verb!*!'' is always placed after ``\verb!%*!'' at the beginning.

\subsection{Predicate Signatures}

\lang\ allows to declare the names and arities of used predicates. Often, predicates play different roles in an encoding, and it can
make sense to explicitly distinguish between these roles. 
In particular, it can make sense to document which are the relations that are defined by the rules of a block (those will
usually appear in the heads of some rules) and which are the relations that are expected to be already defined by some other rules (the required
predicates will usually appear in some rule bodies). This distinction is closely related to  
intensional and extensional predicates in a database setting. 
If a block is regarded as a declarative problem-solving module, the predicates that are required will usually encode problem instances, and respective
predicates form the \emph{input signature} of a block. Accordingly, the relations that are defined form the \emph{output signature} of a block.
We note, however, that input and output signatures might overlap in a declarative setting.

We proceed by declaring the predicate names as well as the input and output signatures of our encoding as described above. 
Within block \verb!Battleship!, we add the following:

{\small
\begin{verbatim}
%** 
 @atom water(X,Y)
 there is no ship at position (X,Y)

 @atom ship(X,Y)
 a ship is occupying position (X,Y)

 @atom rowHint(X,H)
 in row X, H squares are occupied 

 @atom colHint(Y,H)
 in column Y, H squares are occupied

 @atom ship(X1,Y1,X2,Y2)
 a ship is occupying the squares from (X1,Y1) to (X2,Y2)

 @input water/2, ship/2, rowHint/2, colHint/2
        
 @output ship/4 
*%
\end{verbatim}
}

\noindent 
We use \verb!@atom! to introduce the name of a predicate along with its arity and some information describing its
intended use, and we use \verb!@input! and \verb!@output! to determine which predicates symbols are used to represent input for the block
and which ones correspond to output. For input and output signatures, we also have to explicitly give the arity of the involved predicates. 
This is needed to disambiguate between predicates having the same name but a different number of arguments, as \verb!ship! in our running example. 
Note that annotations are optional, declarations are not enforced. 
We stress that declaring input and output signatures should be done only if this is beneficial for the development process and is consistent with
the declarative reading of a program. 
Input and output are terms that are quite commonly used in declarative programming---\egc all problem specifications for the benchmark problems used for
the ASP competitions explicitly define input and output  signatures. However, if rules are seen as more general definitions of relations of some problem domain, 
it might be more appropriate to use \verb!@defines! to declare the relations that are defined by a block of rules and \verb!@requires! to declare
the input signature of a block.
Again, such annotations are not mandatory. 
The more information is made explicit, the more it can be used  by tools
that interpret such information to the benefit of the developer. 

\subsection{Types}

Regarding the declaration of predicates, we can provide information about the types of its term arguments. 
Assume that row and column positions are specified by ascending integers starting from $1$. To make this
assumption explicit, we add the following lines to the block:

{\small
\begin{verbatim}
%** 
 @term X, Y
 @from 1, 2, 3, 4, 5, 6, 7, 8, 9, 10
 X (Y) is a row (column) index ranging from 1 to 10 
*%
\end{verbatim}
}

Here, we use \verb!@term! to declare variable names and \verb!@from! to specify the
type of a variable by means of a non-empty comma-separated list of admissible ground terms.
As an alternative to \verb!@from!, we can use \verb!@with! followed by a rule body:

{\small
\begin{verbatim}
%** @with integer(#V), #V>1, #V<10 *%
integer(1..1000).
\end{verbatim}
}

The type of \verb!X! and \verb!Y! is now determined by the ground substitutions for the 
reserved symbol \verb!#V! that satisfy the rule body given after \verb!@with!. Here, ``\verb!integer(1..1000).!'' is
regular ASP code for defining facts that encode the integer range that we are considering.
Furthermore, to express that variables are of the same type as ones already known, we can use
\verb!@samerangeas!, as illustrated next:

{\small
\begin{verbatim}
%** 
 @term X1, Y1, X2, Y2
 further row and column indices
 @samerangeas X 
*%
\end{verbatim}
}

Thus, any of \verb!X1!, \verb!Y1!, \verb!X2!, and \verb!Y2! is of the same type  as \verb!X!.
For future work, we also plan to extend \lang\ so that predefined names for at least basic types 
like strings or integers can directly be used to specify the types of variables.

As mentioned previously, blocks can be nested. To proceed with our Battleship encoding, we add a block of rules within 
block \verb!Battleship! for guessing solution candidates
according to the usual guess-and-check paradigm:

{\small
\begin{verbatim}
%** 
 @block Guess {
 guess a configuration of battleships on the grid 
*%
r(1..10). c(1..10).
{ship(X1,Y1,X2,Y2):r(X1):c(Y1):r(X2):c(Y2):X2>=X1:Y2>=Y1}.  
:- ship(X1,Y1,X2,Y2), X1!=X2, Y1!=Y2.   
%** } *%
\end{verbatim}
}

Note that in block \verb!Guess!, all declarations for predicates and terms are inherited from the enclosing 
block \verb!Battleship!. One can proceed in a similar way to define blocks for  constraints (along with auxiliary definitions)
to prune away unwanted solution candidates to complete the encoding.

\subsection{Assertions}

Towards testing and the verification of programs, \lang\ allows the formulation of general assertions, as well as pre- and postconditions for blocks.
A precondition is a logical condition that is assumed to hold for any input while a postcondition has to hold on any output of a block.
Together, pre- and postconditions can be regarded as a \emph{contract}: it is the responsibility of any input provider
 that a block's preconditions are satisfied, and it is the responsibility of the rules in the respective block that its postconditions are 
satisfied. Both pre- and postconditions are formulated in ASP itself; they are placed within the block they belong to.
As an illustration, we formulate as a precondition of our Battleship encoding that
 no square shows both water and parts of a ship jointly as follows:

{\small
\begin{verbatim}
%** 
 @precon Excl {
 no square shows both water and a part of a ship
 @never clash
 clash :- water(X,Y), ship(X,Y). 
 } 
*%
\end{verbatim}
}

In general, a precondition is introduced with the keyword \verb!@precon! followed by a name for the condition.
The actual content is then written enclosed between ``\verb!{!'' and ``\verb!}!'', similar to the definition of blocks.
An optional description follows. Then, we have to specify a non-empty list of ground atoms after the keyword \verb!@never! or
\verb!@always!. After that, we add some ASP rules that define the before-mentioned ground atoms.  
The intended semantics is as follows. Some input, \iec a set of facts over a block's input signature, passes a precondition if that input combined
with the rules of the precondition cautiously entails all the ground atoms after \verb!@always! and the negation of  the atoms after \verb!@never!.

Postconditions are expressed analogously to preconditions. To say that battleships must not be longer than four squares, we add
the following code to our Battleship block:

{\small
\begin{verbatim}
%** 
 @postcon Overlength {
 battleships are never longer than four squares
 @never ov
 ov :- ship(X1,Y1,X2,Y2),L=X2-X1,L>4. 
 ov :- ship(X1,Y1,X2,Y2),L=Y2-Y1,L>4. 
 } 
*%
\end{verbatim}
}
\noindent
A block and an input for a block satisfy a postcondition if the block joined with the input and the rules of the postcondition
 cautiously entails all the ground atoms after \verb!@always! and the negation of the atoms after \verb!@never!.

Having pre- and postconditions explicitly formalised offers various ways to support the development process. For one thing,
they can be used to automatically generate  input instances for testing purposes. 
This can be automated for systematic testing
of a block, including random testing and structure-based testing~\cite{janhunen10,janhunen11}. Towards program verification, one can check
whether a block passes its postconditions for any admissible input, at least from some fixed small scope, \iec involving only a bounded number of constant symbols. Exhaustive testing \wrt a
small scope showed to be quite effective in ASP~\cite{oetsch12}. Though they are formulated in ASP itself and
thus tend to duplicate some code from within a block,  pre- and postconditions are especially of great value if the rules in a block are changed, \egc to
optimise an encoding, and one wants to be sure that the changes did not render the program incorrect.
This can be done, e.g., by searching for inputs within some small scope that violate some postcondition of the optimised program, provided respective tool support is available.

We note that pre- and postconditions make only sense in a setting where we can also distinguish between input and output relations. 
If this is not the case and one wants to formulate general assertions on the answer sets of a program,  \lang\ allows to define
them using \verb!@assert!. Semantically, an assert statement is a postcondition of the whole program.

\subsection{Unit Tests}

Though pre- and postconditions allow to partially verify program correctness, \lang\ also supports a light-weight form
of program verification that is inspired by unit testing. A unit test in procedural languages is commonly a test for an individual function or procedure. 
While in a related approach for unit testing answer-set programs~\cite{febbraro11}, the scope of a test is defined in terms of sets of rules,
unit tests are formulated for blocks or sets of blocks in our setting. 
To check whether the guessing part
of our running example generates solution candidates where one ship occupies precisely the first four horizontal squares of the field, we could formulate 
a unit test as follows: 

{\small
\begin{verbatim}
%** 
 @testcase ShipTopLeftCorner
 a ship is horizontally placed at the top-left corner 
 @scope Guess
 @testatoms goalShip @trueinatleast 1 
*%
goalShip :- ship(1,1,1,4).
\end{verbatim}
}

A unit test starts with the reserved word \verb!@testcase! followed by a name. 
Then, a short description of the purpose of the unit test
may be given as a comment. After \verb!@scope!, a list of block names is expected. In the above example, we used
\verb!@testatoms! to declare that \verb!goalShip! is an atom that has to be true in at least one answer set (\verb!@trueinatleast 1!) 
of block \verb!Guess! joined with
the subsequent rule that defines  \verb!goalShip!.
Instead of or additional to \verb!@trueinatleast! $n$, a tester might use 
\verb!@trueinatmost! $m$, \verb!trueinall!, \verb!falseinatleast! $p$, \verb!falseinatmost! $q$, and \verb!falseinall!, where $m$, $n$, $p$, and $q$ are positive integers. 
Also, instead of \verb!@testatoms!, one may use \verb!@testhasanswerset! or \verb!@testnoanswerset! to express that at least one
or no answer set is expected, respectively.

The semantics of a unit test is as follows. A test case passes iff the answer sets of the rules of the test case combined with all the blocks specified after \verb!@scope!
satisfy the testing conditions expressed using any of \verb!@testatoms!,   \verb!@testhasanswerset!, and \verb!@testhasnoanswerset!.
For example, to additionally test that a ship is never placed diagonally on the field, 
one could formulate a further test case:

{\small
\begin{verbatim}
%** 
 @testcase NoDiagonalShips
 ships are never placed diagonally on the field
 @scope Guess
 @testatoms forbiddenShip @falseinall 
*%
forbiddenShip :- ship(1,1,3,3).
\end{verbatim}
}

Of course, this test case can only guarantee that one particular ship is not placed diagonally at some particular position. However, this distinguishes test cases from more general assertions like postconditions. To generalise the above test case to arbitrary ships, we would rather use a postcondition. Typically, test cases represent concrete situations by means of facts that can be easily verified by a user and document individual situations that are allowed or forbidden. They cannot, however, guarantee correctness of an encoding but only increase our confidence regarding its functionality.

Unit testing is a convenient way to test properties of individual blocks of an ASP encoding. Furthermore, they can be used to iteratively develop programs in 
a test-driven fashion. In test-driven development, unit tests are formulated before the code is written. First, a unit test for a single property of the block
that we want to develop is specified. Then, it is checked whether the test case fails for the program under development. If this is the case, the block is extended by
the necessary rules 
 to make the failed test case pass. After the code is refactored towards efficiency, readability, etc., and after it is verified that all test cases still pass, 
the next property is addressed by formulating a respective unit test. This continues until the program is complete. 
 For illustration, the unit tests \verb!ShipTopLeftCorner!  and \verb!NoDiagonalShips! will pass for the current state of the Battleship program. 
 However, if we want to implement the property that ships must not touch each other, we could specify the following test case (which will currently fail):

{\small
\begin{verbatim}
%** 
 @testcase TouchingShips
 two ships must not touch each other
 @scope Guess,Touch
 @testatoms forbiddenShip @falseinall
*%
forbiddenShip :- ship(1,1,1,2), ship(1,2,1,4).
\end{verbatim}
}
\noindent
Then, we would proceed to implement a block \verb!Touch! with a constraint that
forbids answer sets with ships that are touching each other and check whether the new 
test case and the old ones pass.

\section{\aspdocBf}\label{sec:aspdoc}
\aspdoc is a command-line tool that interprets meta-information given in an answer-set program and generates a corresponding HTML documentation file similar to, \egc\ \javadoc for Java programs. Information regarding block structure, input and output signatures, used predicate symbols, etc. is clearly arranged so that
the answer-set program can easily be understood, used,  or extended by other developers. 
Such documentation features are especially useful to make ASP problem-solving libraries, \iec collections of ASP encodings that can be used as 
building blocks for larger programs, accessible to developers.

The tool is developed in Java; an executable JAR file is available on the web.\footnote{%
\url{http://students.sabanciuniv.edu/dgkisa/aspdoc-aspunit}.}
Assume that the  source code of our running example is stored in a file, say \verb!battleship.gr!.  A corresponding HTML documentation can be created as follows:

\begin{verbatim}
   java -jar aspdoc.jar -p battleship.gr
\end{verbatim}

\noindent
Different HTML documents are created with \verb!index.html! as the usual entry point.
Here, option \verb!-p!, or \verb!-potassco!, is used to tell \aspdoc that the answer-set program is written using \gringo syntax. 
For \DLV, option \verb!-d! or \verb!-dlv! can be used instead. 
Furthermore, an output directory $d$ can be specified with option \verb!-o=!$d$, and help on available options can be obtained with the option \verb!-h!.
A summary of \aspdoc options is given in Table~\ref{tab:aspdoc}.

\begin{table}[t]
\caption{Command-line options for \aspdoc.}\label{tab:aspdoc}
\centering
\begin{tabular}{ll}
\hline\hline
Option		&	Description \\
\hline
\verb!-o=!\emph{path} 	& set output directory to \emph{path} \\
\verb!-HA! 			& show hidden atoms \\
\verb!-ha! 				& do not show hidden atoms \\
\verb!-S! 				& include ASP code in the HTML document \\
\verb!-s! 				& do not include ASP code in the HTML document \\
\verb!-A!				& include \lang\ code in generated ASP code \\
\verb!-a!				& do not include \lang\ code in generated ASP code \\
\verb!-potassco!, \verb!-p! 	& input language is that of \gringo \\
\verb!-dlv!, \verb!-d! 		& input language is that of \DLV \\
\verb!-help!, \verb!-h!		& print usage information \\
\hline\hline
\end{tabular}
\end{table}

\begin{figure*}[t]
\figrule
\vspace{4pt}
\includegraphics[width=\textwidth]{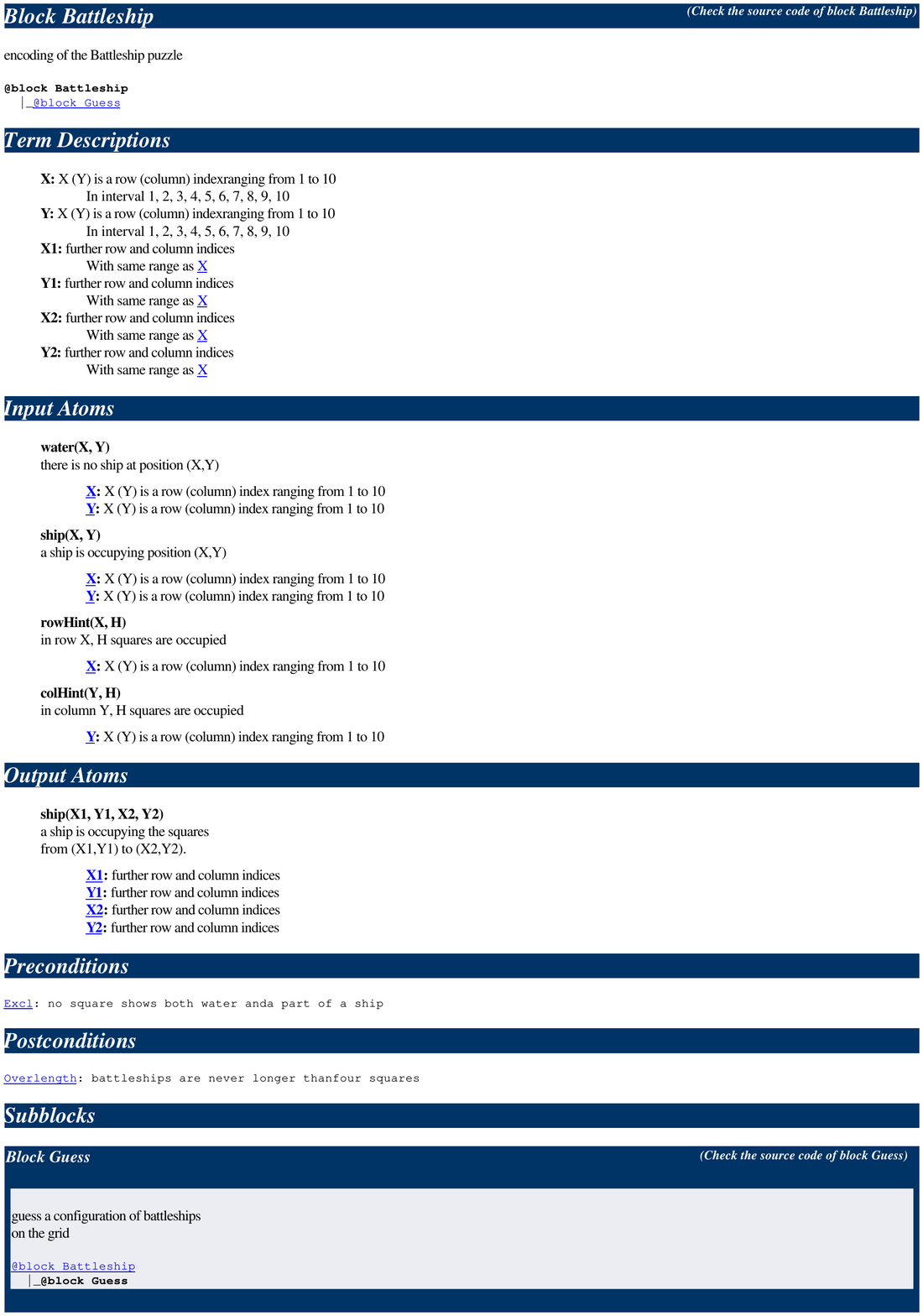}
\vspace{-7pt}
\figrule
\caption{HTML documentation of the Battleship program.}\label{fig:screenshot}
\end{figure*}

A screenshot of the documentation  for the Battleship example presented above is given in Fig.~\ref{fig:screenshot}.
The documentation of the complete encoding can be found online.\footnote{\url{http://www.kr.tuwien.ac.at/research/projects/mmdasp/battleship}.}
The document contains 
descriptions of all the blocks of the answer-set program, where sub blocks are indented relative to their parent blocks.
To provide an overview, a summary of the block structure of the entire answer-set program is presented at the beginning of the documentation.
We note  that programmers are not forced to declare blocks at all.
If no
block is specified in  a file, all rules in that file belong to a dedicated default block.
For each block, 
descriptions of the used predicates and types of involved terms, as well as pre- and postconditions are given.
By default, hidden atoms, \iec atoms mentioned  neither in a block's input nor in its output signature, are displayed as well  in a dedicated section entitled ``Hidden Atoms''. To hide them,
option \verb!-ha! can be used. 
The document also contains a link to the actual rules inside a block, unless this is suppressed using option \verb!-s!. These rules are, by default, displayed together
with the meta-comments of \lang. If option \verb!-a! is used, such comments are not shown.
Likewise, the rules for defining pre- and postconditions can be inspected by using the respective links. 
To enhance navigability between different parts of the document, predicates used in the source code view or in signature declarations are, if available, linked to their respective descriptions. For instance, to find out the range of a variable in the output atoms section, say $X1$, the user can simply follow the link and thereby navigate to the description of $X1$ (and $X$) in the term description section. Further options to customise the appearance of the documentation are planned for future work.

\section{\aspunitBf}\label{sec:aspunit}

\aspunit is a  tool to execute test cases that are formulated in \lang.  Like \aspdoc, 
it is
a command-line tool. An executable JAR file can be downloaded from the same web page as \aspdoc.
For \aspunit, each unit test has to be stored in a separate file. Although test cases are required to have a name, we allow that a user may omit an explicit name, in which case the file name is used by default.
The tool takes as input a test-suite specification file, \iec a file that contains the relevant information regarding locations of the answer-set program, the files containing individual
test cases, and the ASP solver that is needed to execute them.
The syntax of a test-suite specification file
 is closely related to our annotation language itself. In particular, the specification of a test-suite has the following form:

{\small
\begin{verbatim}
@testsuite name
 description
@program    ASPfiles
@programdir pathToASPfile
@test       testCaseFile1
@test       testCaseFile2
 ...
@testdir    pathToTestFiles
@solvertype ASPsolver
@solver     solverFile
@grounder   grounderFile
\end{verbatim}
}
\noindent
Hence, a test-suite specification starts with \verb!@testsuite! followed by a name. Then, a short description may be given.
A list of  file names that  together contain the answer-set program under test is expected after \verb!@program!. These file names are relative to a path specified after
\verb!@programdir!. For each test case that we want to execute, we have to provide the file name that contains that
test case specified by \verb!@test!. The path to these files appears after \verb!@testdir!. Then, information regarding the ASP solver has to be given. 
For this, \verb!@solvertype! is used; the solver type  is one of 
\DLV, \clasp, or \clingo. After \verb!@solver!, an absolute file name of the ASP solver is expected. This file name may include additional parameters for that solver. 
If a separate grounder is needed, like for \clasp, an absolute file name including command-line parameters have to be specified after \verb!@grounder!.

Now, to run a bunch of test cases specified within a test-suite file, say \verb!testsuite!, \aspunit is invoked as follows:

\begin{verbatim}
   java -jar aspunit.jar testsuite
\end{verbatim}

\noindent
The tool will run all the unit tests on the 
answer-set program using the solver settings according to specifications in  \verb!testsuite!. A test report is printed to standard-output. 
This report contains information regarding success or failure for each test case. If a test case fails, a counterexample may be included, depending whether
option \verb!-CE! is set when \aspunit is executed. Furthermore, if option \verb!-D! is used, the test report will contain a short description of each test case that fails, obtained from the specification of
the test cases themselves.
For illustration, assume we run the test cases presented in Section~\ref{sec:lang} on the partial encoding of Battleship. 
Recall that the first and second test case pass while the third one fails. The resulting test report, including a description of each test case  and counterexamples for the failing test case, is given in Fig.~\ref{fig:testreport}.
A summary of \aspunit command-line options is given in Table~\ref{tab:aspunit}.

\begin{figure}[t]
\centering
\figrule
\small
\begin{verbatim}
Test Case ShipTopLeftCorner: Successful

Test Case NoDiagonalShips  : Successful

Test Case TouchingShips    : Failed
    two ships must not touch each other
  
  Failed Test : @falseinall forbiddenShip
    Counterexample:
      Answer set:
        c(1).c(10).c(2).c(3).c(4).c(5).
        c(6).c(7).c(8).c(9).forbiddenShip.
        r(1).r(10).r(2).r(3).r(4).
        r(5).r(6).r(7).r(8).r(9).
        ship(1, 1, 1, 2).ship(1, 2, 1, 4).
\end{verbatim}
\figrule
\caption{A test report for the Battleship program.}\label{fig:testreport}
\end{figure}

\section{Related Work}\label{sec:rel}

As mentioned earlier, there are many notions of modularity for ASP in existence~\cite{bugliesi94,eiter97,gelfond99,balduccini07,janhunen09}.
The advantage of the light-weight approach of \lang\ for grouping rules together by means of declaring  blocks is that we do not change
the semantics of programs and they can be directly parsed by ASP solvers.
However, there are also disadvantages 
compared to other approaches to modularise logic programs. 
For example,  related to our approach for annotating type information is {\sc RSig}~\cite{balduccini07}. {\sc RSig} is 
an language extension for specifying simple type information for programs and modules and thus requires its own parser. However, 
this type information simplifies the program rules as information about the type of variables is not required to be provided.
On the other hand, DLP-functions~\cite{janhunen09} allow to compute the semantics of a logic program based on the semantics of its
separate DLP-functions which opens up new paths for potentially more efficient answer-set computation. 
To support the incremental development of logic programs, $\mathtt{lp}$-functions can be used to structure a program and to develop it
along with its specification by using specification constructors and their realisation theorems~\cite{gelfond99}. Blocks in \lang\ are not designed
to serve one particular purpose, different interpretations  are conceivable and are eventually determined by respective tool support like \aspunit
for unit testing blocks of rules.

In general, developing and debugging a declarative language is quite different from software engineering in a more traditional procedural or object-oriented programming language. With larger programs for real-world applications being written, it is vital to support the programmer with the right tools.
In recent years, some work has been done to provide the ASP programmer with dedicated tools.
The integrated development environments \ape~\cite{sdvbf07} and \sealion \cite{sealion} provide, among other features, syntax colouring  and syntax checking for ASP programs and run as an Eclipse front-end to solvers.
IDEs for the \DLV solver and its extensions are discussed by \citeN{peritecive07} and \citeN{aspide}. 
Debugging in ASP is supported by \textsc{spock} \cite{brgepusctowo07b}, which makes use of ASP to explain and handle unexpected outcomes like missing atoms in an answer set or the absence of an answer set. 
\citeN{aspviz} and \citeN{kara} provide mechanisms to visualise answer sets of a given program 
to support code debugging.

\begin{table}[t]
\caption{Command-line options for \aspunit.}\label{tab:aspunit}
\centering
\begin{tabular}{ll}
\hline\hline
Option		&	Description \\
\hline
\verb!-CE!  			& show counterexample if a test case fails \\
\verb!-ce!  				& do not show counterexample if a test case fails \\
\verb!-D!				& show description of a test case if it fails \\
\verb!-d!				& do not show description of a test case if it fails \\
\verb!-help!, \verb!-h!		& print usage information \\
\hline\hline
\end{tabular}
\end{table}

To support large application developments, traditional languages offer programming tools that automatically generate searchable documentation, like \egc\ \javadoc.
Methodologies like test-driven development provide 
a mechanism
 to incrementally unit test   code and to support regression testing; \junit
is an example of this for  Java.
 \lang\ provides the support for both, incorporating the annotation of tests directly into the documentation of the program. 
The use of assertions in \lang\ is inspired by the Java Modelling Language \cite{Leavens06designby} and annotations as used in Prolog \cite{kulas00}. 

Similar to our unit testing approach,  Prolog offers unit-testing functionality called \plunit.\footnote{\url{http://www.swi-prolog.org/pldoc/package/plunit.html}.}
As in our approach, where  tests are expressed using ASP itself and only a Java wrapper is used to call all tests within a given test-suite, tests in 
\plunit\ are formulated as Prolog clauses.

\citeN{febbraro11} provide a mechanism for unit testing in ASP, which is incorporated in their IDE \aspide. They base unit tests on clusters on the dependency graphs or rule labelling, while we allow the user to decide which rules belong to a test by defining blocks.

\section{Conclusion and Future Work}\label{sec:concl}

In this paper, we presented \lang, an annotation language for ASP. This language can be used to structure a program into blocks and to
declare language elements like predicates with type information, input and output signatures, pre- and postconditions, test cases, etc.
Annotations do not interfere with the languages of answer-set solvers as they have the form of program comments. 
The main advantage of such annotations is that they can be interpreted by tools to support the development process, to automatically
test and verify programs, and to increase maintainability by enhancing program documentation.
In fact, we implemented and described two such tools, namely \aspdoc for generating an HTML documentation for a program, and \aspunit
for running and monitoring unit tests. The former tool is especially useful for maintaining and using larger collections of program modules; the latter
tool is used for managing a test corpus when a program is developed and to enable test-driven development methods, a methodology popular in industry, \egc as in extreme
and agile programming.

While many interesting features of \lang\ for development support can be realised by stand-alone tools like \aspdoc and \aspunit, things become more interesting when the respective functionalities are available within an IDE for ASP. The two most actively developed IDEs for ASP at present are \sealion~\cite{sealion}
and \aspide~\cite{aspide}. Then, the proposed language can be used as a basis to realise intelligent syntax highlighting, static or dynamic type checking, code completion, and so on. 
Indeed, \lang\ can already be parsed by \sealion and the features of ASPDOC are already available from within the IDE.
Further integration is planned with the goal that all features sketched in this paper are supported in \sealion.

Furthermore, we want to empirically evaluate
to what extent additional meta-information is beneficial for program development within courses on declarative problem solving at our universities.
In general, program annotations provide a wealth of information. One of the main issues with debugging ASP programs is the difficulty of working out the program's interpretation of the problem (resp., solution), and the programmer's view of the problem (resp., solution). Using the
meta-data, it would be possible to automatically generate a semi-natural language reading of the program, allowing programmers to cross-check their interpretation of the program with that of the program itself.
This is only possible if developers use a specific grammar to annotate the various components of the program. 

In traditional software engineering, coding standards including documentation are imposed, especially in case of developing large software projects. In this paper, we propose \lang\ as part of coding standards for ASP. Future work will look into other best-practices for  writing and maintaining ASP programs. The growing number of applications of ASP can provide a wealth of information for this.

\label{lastpage}

\end{document}